\documentclass[12pt]{article}
\usepackage{amssymb}
\usepackage{amstext}
\newcommand*\de{\mathrm{d}}
\newcommand*\De{\mathrm{D}}
\renewcommand*\epsilon{\varepsilon}
\renewcommand*\phi{\varphi}
\renewcommand*\theta{\vartheta}
\begin{document}

\title{Mathisson-Papapetrou equations in metric and gauge 
theories of gravity in a Lagrangian formulation}
\author{ M. Leclerc}
\maketitle 
\begin{abstract}
We present a simple method  
to derive the semiclassical equations of motion for 
a spinning particle in a gravitational field. 
We investigate the cases of classical, rotating  
particles, i.e. the so-called pole-dipole particles, as well as 
particles with an additional intrinsic spin. We show that, starting 
with a simple Lagrangian, one can derive equations for the spin evolution 
and momentum propagation in the framework of metric theories of gravity 
(general relativity) and in theories based on a  Riemann-Cartan geometry 
(Poincar\'e gauge theory), without explicitly 
referring to matter current densities (spin and stress-energy). 
Our results agree with those derived from the multipole expansion of 
the current densities by the conventional Papapetrou method and 
from the WKB analysis for elementary particles. \\[0.5cm] 
\end{abstract}

\section{Introduction}

It is of interest to have an alternative method of deriving the 
Mathisson-Papapetrou equations (see \cite{1} and \cite{2}) not only 
because the conventional procedure 
is rather lengthy, but more importantly, because one might get a better 
insight into the interpretation of the various quantities involved
(see also \cite{3} for a detailed analysis of the classical method). 
For instance,  in 
Riemann-Cartan space, there exists a large variation of  
possibilities of  how to define the momentum vector which eventually  lead
 to differences in 
the final equations  (see \cite{4} and \cite{5}, 
especially the remarks at the end of \cite{5}). Furthermore, if we 
consider a specific solution for the metric and the torsion fields, 
it would be convenient to have a Lagrangian into which we can plug 
 the results and derive the equations of motion by varying the Lagrangian, 
instead of plugging the solution directly into the equations of motion. 
For instance, in general relativity, one usually substitutes the 
Schwarzschild metric into the Lagrangian $L = mg_{ik}u^iu^k$ and not into 
the geodesic equation, which would require to evaluate the complete 
Christoffel symbols. 
In this way, one can take full advantage of the symmetries 
of the problem. The aim of this article is therefore to find 
suitable generalizations of $L=mg_{ik}u^iu^k$
that allow  for the description of particles with spin (classical and 
intrinsic) moving in a Riemann-Cartan geometry. 

To avoid confusion, let us make a remark on the language we will 
use throughout this article. We will use the expression \textit{classical 
spin} of a body to denote its orbital angular momentum in its (center 
of mass) rest frame. The word classical is used to distinguish it 
from the \textit{intrinsic spin} which is of quantum mechanical origin, 
and the word spin is used to emphasize that we are always talking about 
the angular momentum in the rest frame. When talking about \textit{intrinsic 
spin}, we either refer to the intrinsic spin of an elementary particle, 
or to the total intrinsic spin of a macroscopic body as a result of 
the polarization of its constituent particles, for instance in a ferromagnet 
or in a neutron star. The word spin may refer to both classical and 
intrinsic spin.

Several attempts have been made for a Lagrangian description in the past and 
results were achieved   not only  
in general relativity for the classical multipole particle \cite{6}, 
but also for point particles with intrinsic spin  
in a Riemannian space \cite{7}, \cite{8}. We will 
review these cases, discuss differences in our 
derivation method and 
 generalize to  particles with intrinsic spin and 
 dipole moment (classical spin) in a general Riemann-Cartan space. 

Let us use the example of a test particle 
in an electromagnetic field to illustrate 
how we will manage to achieve 
equations for the position as well as for the spin evolution, 
from a single Lagrangian, 
without 
referring to the field equations or the Bianchi identities.   

The Lagrangian of a charged particle  reads
\begin{displaymath}
L = \frac{m}{2} g_{ik} u^iu^k - e A_i u^i. 
\end{displaymath}
The charge $e$ is of course the integrated current $j^m$ and its 
conservation law $\dot e = 0$ (the dot denotes the time derivative 
with respect to proper time $\tau$) can be derived from the relation
$j^m_{\ ;m} = 0$ which is a consequence of the field equations. 
However, we do not want to refer to the field equations at all and we will 
ignore 
the relation $\dot e = 0$ for the moment. 
The equations of motions are then derived in the form 
\begin{displaymath}
m \hat \De u^i = e F^i_{\ k} u^k  + A^i \frac{\de e}{\de \tau}. 
\end{displaymath}
The symbol $\hat \De$ denotes the covariant derivative with respect to the 
proper time $\tau$. 
Without knowing the 
Maxwell equations, we certainly know that they are   
gauge invariant with respect to $A_m \rightarrow A_m + f_{,m}$ 
 and 
consequently 
this should be the case for the equations of motion too. 
We therefore have to require $\dot e = 0$. In this way, we 
are let to the charge conservation equation as well as to the 
correct equation of motion. How do we know that 
the field equations will be gauge invariant? Well, the action for 
the Maxwell field is constructed under exactly this requirement. 
The gauge invariance is more fundamental then the field equations 
themselves. It incidentally leads to the unique Lagrangian 
$F^2$. (This will not be the case for the local $SO(3,1)$ gauge 
invariance of Poincar\'e gauge theory, which allows for a whole 
family of Lagrangians.) 

In the same way, we will derive the equations for the evolution of 
the intrinsic spin 
and/or classical spin by investigating the covariance properties 
of the momentum equation with respect to coordinate transformations 
and/or Lorentz gauge transformations. 
 
In the next section, we  derive in this way the Mathisson-Papapetrou equations 
for the pole-dipole particle in general relativity. In section 3, we 
generalize the result to Riemann-Cartan geometry and finally in 
sections 4 and 5 we investigate the case of particles with intrinsic spin
and of macroscopic spin polarized bodies. To get more confidence in 
our procedure, we will re-derive in section 6 the equations of motion 
for the 
case of the Dirac particle using an alternative method. In section 
7 we consider as an illustrative example the precession equation of 
a spinning particle in the torsion field arising in Brans-Dicke theory 
generalized to Riemann-Cartan space. Finally, in section 8, we analyze 
the symmetry properties of the Lagrangian itself, as opposed to 
the equations of motion.

\section{The classical Mathisson-Papapetrou equations}

In this section, we consider a classical (i.e. without intrinsic spin) 
pole-dipole particle moving 
in a curved, Riemannian spacetime. 

Instead of postulating a Lagrangian, we will try to 
derive it from simple considerations. The following manipulations,  
although  not  mathematically rigorous, will provide us with a 
strong motivation for the Lagrangian (8) which will be our starting point. 

In general relativity, the geodesic equations for a body of mass $m$ 
are derived from the following 
Lagrangian
\begin{displaymath}
L = \frac{m}{2} g_{ik} u^iu^k. 
\end{displaymath}
First, consider the body to be constituted from freely moving test 
particles, moving all in a given exterior field. 
Then, the Lagrangian can be written as
\begin{equation}
L = \sum_l \frac{m_{(l)}}{2} g_{ik}(x_{(l)}) u^i_{(l)}u^k_{(l)}.  
\end{equation}

However, in order to describe an extended, rotating body, 
the coordinates of 
the different constituents will have to fulfill certain constraints, 
due to the rigid structure of the body. 
We now follow the method that is usually used on post-Newtonian 
Lagrangians (see \cite{9} for instance) and also  in 
classical mechanics,  but 
we will apply it directly to the covariant form (1). Let $X^i$ denote the 
center of mass of the body. (We avoid the difficulties and ambiguities 
concerning the definition of the center of mass in general relativity. 
We just suppose that there exists a vector $X^i$ whose spatial 
components $X^{\alpha}$, $\alpha = 1,2,3$ agree in some limit with 
the Newtonian center of mass.)  

Then, for the coordinate of a mass element 
of the body we can write (we omit the index denoting the mass element) 
\begin{equation}
x^i = X^i + \rho^i, 
\end{equation}
and for the velocity of the same mass element  
\begin{equation}
u^i = V^i + \omega^i_{\ k} \rho^k,  
\end{equation}
with antisymmetric $\omega^{ik}$ and $V^i = \de X^i / \de \tau$. 
Just as in the original work of Papapetrou
 \cite{2}, 
we suppose $\rho^0 = 0$ in a certain reference frame.  
If we set in the same frame $\omega^{i0}= 0$, 
the quantity $\omega^{ik}$ is clearly the 
angular velocity in four dimensional form, i.e. we have for the 
spatial part $\epsilon^{\nu \mu \lambda} \omega_{\lambda} 
=  \omega^{\mu \nu}$ and the 
three velocity reads $\vec u = \vec V + \vec \omega \times
\vec \rho$. Consequently, $\omega^{ik}$ is related to the angular momentum 
(the classical spin) through 
\begin{equation}
S^{ik} = I \omega^{ik}, 
\end{equation}
where $I$ is the moment of inertia of the body, defined as 
in classical mechanics, i.e. we have in the spherical 
 case $I = I^{xx} = I^{yy} = I^{zz}$, with 
$I^{\mu\nu} =  \int{(-g^{\mu\nu}\rho^2 - \rho^{\mu}\rho^{\nu})}\de m $.  

We will use the three dimensional relation 
\begin{equation}
\frac{1}{2} I g^{\mu\nu} = -\int \rho^{\mu} \rho^{\nu}\ \de m.
\end{equation}
This relation can be found in \cite{10} (\S 106, problem 4) and 
also in \cite{9}. It is 
easily proved by taking its trace and using the fact that $I^{xx} 
= I^{yy} = I^{zz}$. (In the expression for the moment of inertia, the 
metric can be considered to be independent or $\rho$, since the integrant is 
already of order $\rho^2$. See expansion below.) Note that (5) is essentially 
a Newtonian order relation. There is no univoque generalization of the concept 
of moment of inertia in the framework of general relativity.

Since by definition, $X^i$ is the particle's center of mass, we have 
\begin{equation}
\int \rho^i \de m = 0. 
\end{equation}
This equation, again, can only be correctly understood at the Newtonian 
level. We do not specify the integration meter $\de m$ in general, but 
just assume 
that it has the correct Newtonian limit. 
We now write (1) in the form $L = \int \frac{1}{2} g_{ik}(x) u^iu^k \ \de m$ 
and expand the metric $g_{ik}(x)= g_{ik}(X) + g_{ik,l}(X)\rho^l$. 
By using (3)-(6), and  retaining only the first order metric perturbations 
and terms at most linear in $\omega^{ik}$ we find   
\begin{equation}
L = \frac{m}{2} g_{ik} V^iV^k  -
\frac{1}{2}g_{ik,m}S^{im}V^k,  
\end{equation}
 where $V^i = \de X^i / \de \tau$. Note that the terms linear in $\rho$ 
do not contribute because of (6). Since $\rho^i$ has been eliminated, 
we can now leave the frame where $\rho^0 = \omega^{i0}= 0$ and consider 
$S^{im}$ as a four dimensional antisymmetric tensor.   

Using the symmetry properties of the Christoffel symbols $\hat \Gamma^l_{km}$ 
and of $S^{im}$, 
we can finally write (7) 
in the simple form (we change the notation from $V^i$ to $u^i$)
\begin{equation}
L = \frac{m}{2} g_{ik}u^iu^k - \frac{1}{2} \hat \Gamma^l_{km}S_{l}^{\ k} u^m. 
\end{equation}
This Lagrangian (generalized to higher order multipole terms)
  has been used in \cite{6}, starting from Newtonian quantities and 
looking for four dimensional generalizations. A Lagrangian approach 
for spinning media can also be found in \cite{11}. 

Let us emphasize once again that the whole dervivation that led to 
(8) is essentially based on Newtonian physics and can therefore not 
be used to conclude, or even to prove,  that (8) is the correct Lagrangian 
for the description of the dipole particle. The arguments make however 
clear that (8) is probably a good candidate to start with, 
because by construction,  
it will certainly  lead to equations that possess the correct Newtonian 
limit. Beyond that, we can only compare the equations of motion dervived 
from (8) with those derived by other means (as the Papapetrou method) 
or directly with experiment. 
In this sense, we simply postulate the 
Lagrangian (8) and  forget about its origin.

Since $S^{ik}$ is an integrated quantity, it is a 
function of proper time only, without explicit $x^i$-dependence.   
Therefore, the Euler-Lagrange equations are found to be 
\begin{equation}
m \hat \De u_m = - \frac{1}{2} (\hat\Gamma^l_{ki,m}- 
\hat\Gamma^l_{km,i})S_l^{\ k}u^i + \frac{1}{2}\hat \Gamma^l_{km}
\dot S_l^{\ k}.
\end{equation}
 
Next, the authors of \cite{6}  proceed as follows. In a  coordinate system 
 where $\hat\Gamma^l_{ik} = 0$,  equation (9) can be written as  
\begin{equation}
m \hat \De u_m = - \frac{1}{2} \hat R^l_{\ kmi}S_l^{\ k}u^i, 
\end{equation}
and since this is a tensor equation, it has to be valid in every 
coordinate system. In this way, starting with a more general Lagrangian, 
they were able to derive 
 equations of motion very effectively, including up to octupole moment 
contributions. Apart from (10), the spin evolution equation was 
written down as covariant generalization of the Newtonian precession 
equation. 
In brief, this  method consists in looking for covariant equations that 
have the correct Newtonian limit. In a Riemannian theory, this is 
a very powerful tool and is infallible in the sense that there is only one 
covariant generalization to a given Newtonian equation, as long as we 
suppose that the minimal coupling prescription is valid (that is, matter 
fields should not couple directly to curvature). 

Here, we will follow a different procedure. 
Since we will later on generalize our geometry 
to Riemann-Cartan space, there will in general be more than one way 
to write a certain Newtonian equation in a covariant form. 
For instance, both the geodesic and the autoparallel equation, although 
they differ in a general Riemann-Cartan geometry and although they are 
both covariant,  
reduce to the correct, free particle motion in the absence of 
gravitational fields (curvature and torsion). Therefore, the mere knowledge 
of the Newtonian limit is not enough to fix the general 
form of an equation. 

We  
proceed as in the electromagnetic analogy in the introduction. We claim 
that (9) is already covariant as it stands and look at the consequences 
of this claim. If we complete the first two terms of the r.h.s. of 
(9) to form the curvature tensor, we find 
\begin{eqnarray*}
m \hat \De u_m &=& - \frac{1}{2} (\hat\Gamma^l_{ki,m}- 
\hat\Gamma^l_{km,i} + \hat \Gamma^l_{nm}\hat \Gamma^n_{ki} 
- \hat \Gamma^l_{ni}\hat \Gamma^n_{km} )S_l^{\ k}u^i \\
&&+ \frac{1}{2}( \hat \Gamma^l_{nm}\hat \Gamma^n_{ki} 
- \hat \Gamma^l_{ni}\hat \Gamma^n_{km} )S_l^{\ k}u^i
+ \frac{1}{2}\hat \Gamma^l_{km}
\dot S_l^{\ k}.
\end{eqnarray*}
It is now easy to recognize the covariant derivative $\hat \De S_l^{\ k}$ 
in the second row. Thus, we have 
\begin{equation}
m \hat \De u_m =  - \frac{1}{2} \hat R^l_{\ kmi}S_l^{\ k}u^i 
+ \frac{1}{2}\hat \Gamma^l_{km} \hat \De S_l^{\ k}.
\end{equation}
For this equation to be covariant, we have to require in addition 
\begin{equation}
\hat \De S^{ik} = 0. 
\end{equation}
In this way, we finally get  equations (10) and (12) from the Lagrangian 
(8). They are in agreement with the Mathisson-Papapetrou 
equations in the required 
order of precision, i.e. if one identifies momentum with $mu^i$ (see 
below). 

To summarize our procedure, we suppose that  equations (11) are 
correct (apart from higher multipole terms of course) and since 
ultimately, they were derived from (1), they have to be covariant. 
This claim leads to a constraint, the spin evolution equation.  
There is a certain similarity between this method and  the original derivation 
of Papapetrou  \cite{2}. Papapetrou  begins with a symmetric 
stress-energy tensor and finds, after having expanded up to the 
dipole order, an expression for the integrated stress-energy tensor 
that is not apparently symmetric. The claim that it should be symmetric 
then leads to a constraint which is just the spin evolution equation. 

Last, we remark that there is a tiny difference between (10) and (12) and 
the equations originally 
derived by Papapetrou in the sense that the momentum $P^i$
of the latter is identified with $mu^i$ in our equations, whereas 
the correct relation reads $P_i = mu_i + \hat \De S_{ik} u^k$. 
We will derive the correct equations in the next section; we note however 
that the difference can be shown to be of second order in the spin 
and of first order in the curvature (see \cite{3} or \cite{12})
and is thus negligibly small compared to the other terms.
On the other hand, it has been shown in \cite{13} that (10) and (12) can 
actually be derived from the original Papapetrou equations by a 
slight redefinition of the center of mass of the body and can thus 
 equally well be considered as exact (since we have not specified 
our definition of center of mass). (See also \cite{9} on the 
issue of the center of mass redefinition.)  

\section{The pole-dipole particle in Riemann-Cartan space}

We now turn to a classical pole-dipole particle moving in Riemann-Cartan 
geometry. We know that, since torsion couples only to the intrinsic 
spin, the classical pole-dipole particle should behave exactly  in 
the same way as in a purely Riemannian space. It is however necessary 
to have at our disposal a consistent method of re-deriving the 
same equations in the full Riemann-Cartan framework for the later 
generalization to particles with intrinsic spin. 

Let us first give a short  review of 
the basic concepts of Riemann-Cartan geometry and
fix our notations and conventions. For a complete introduction 
into the subject, the reader may consult \cite{14}-\cite{17} and 
the extensive reference list in \cite{17}.

Latin letters from 
the beginning of the alphabet ($a,b,c\dots $) run from 0 to 3 and 
are (flat) tangent space indices. Especially, $\eta_{ab}$ is the 
Minkowski metric $diag(1,-1,-1,-1)$ in tangent space. Latin letters 
from the middle of the alphabet ($i,j,k \dots $) are indices in a curved 
spacetime with metric $g_{ik}$ as before. We introduce the independent gauge 
fields, the tetrad  $e^a_m $ and the connection 
$\Gamma^{ab}_{\ \ m}$ (antisymmetric in $ab$), as well as 
the corresponding field 
strengths, the curvature and torsion tensors
\begin{eqnarray*}
R^{ab}_{\ \ lm} &=& \Gamma^{ab}_{\ \ m,l} - \Gamma^{ab}_{\ \ l,m} 
                     + \Gamma^a_{\ cl}\Gamma^{cb}_{\ \ m}   
- \Gamma^a_{\ cm}\Gamma^{cb}_{\ \ l}  \\ 
T^a_{\ lm} &=& e^a_{m,l} - e^a_{l,m} + e^b_m \Gamma^a_{\ bl}
- e^b_l \Gamma^a_{\ bm}. 
\end{eqnarray*}
The spacetime connection $\Gamma^i_{ml} $ and the spacetime metric $g_{ik}$
can now be defined through 
\begin{eqnarray*}
e^a_{m,l} + \Gamma^a_{\ bl} e^b_m &=& e^a_i \Gamma^i_{ml}\\
e^a_ie^b_k \eta_{ab} &=& g_{ik}.  
\end{eqnarray*}
It is understood that there exists an inverse to the tetrad, such that 
$e^a_i e_b^i = \delta^a_b $. It can easily be shown that the connection 
splits into two parts, 
\begin{displaymath}
\Gamma^{ab}_{\ \ m} = \hat \Gamma^{ab}_{\ \ m} +
K^{ab}_{\ \ m},
\end{displaymath} 
such that $\hat \Gamma^{ab}_{\ \ m}$ is torsion-free 
and is essentially a function of $e^a_m$. $K^{ab}_{\ \ m}$ is 
 the contortion tensor (see below). Especially, 
the spacetime connection $\hat \Gamma^i_{ml}$ constructed from 
\begin{displaymath}
e^a_{m,l} + \hat \Gamma^a_{\ bl} e^b_m = e^a_i \hat \Gamma^i_{ml}
\end{displaymath}
is just the (symmetric) 
Christoffel connection of general relativity, a function of 
the metric only. This notation is consistent with that of the previous
section.

The gauge fields $e^a_m $ and $\Gamma^{ab}_{\ \ m} $ are 
covector fields with respect to the spacetime index $m$. Under a local 
Lorentz transformation in tangent space, $\Lambda^a_{\ b}(x^m)$, 
they transform as 
\begin{equation}
e^a_m \rightarrow \Lambda^a_{\ b}e^b_m, \ \ \ \Gamma^a_{\ bm} \rightarrow 
\Lambda^a_{\ c}\Lambda^{\ d}_{b} \Gamma^c_{\ dm} - 
\Lambda^a_{\ c,m} \Lambda^{\ c}_{b}. 
\end{equation}
The torsion and curvature are Lorentz tensors with respect to their 
tangent space indices as is easily shown. The 
contortion $K^{ab}_{\ \ m}$ is also a Lorentz tensor and is related to 
the torsion through $K^i_{\ lm}= \frac{1}{2}(T_{l\ m}^{\ i}+ T_{m\ l}^{\ i}
- T^i_{\ \ lm})$, 
with $K^i_{\ lm} = e^i_a e^{}_{lb} K^{ab}_{\ \ m}$ and analogously 
for $T^i_{ \ lm}$. The inverse relation is $T^i_{\ lm} = 
-2 K^i_{\ [lm]}$. 
   
Although we introduce the torsion tensor, 
which can be interpreted as the field strength corresponding to the 
translational group, 
we do not consider (local) 
translations here. 
We require our theory to be 
invariant with respect to  local (tangent space) 
Lorentz transformations and general 
coordinate (spacetime) transformations only. The identification of 
the tetrad (or its nontrivial part) 
as translational gauge potential needs a special treatment and is 
not needed here. (See however section 8.) 

All quantities constructed from the torsion-free connection 
$\hat \Gamma^{ab}_{\ \  m}$ or $\hat \Gamma^i_{lm}$ will be denoted with 
a hat, for instance $\hat R^{il}_{\ \ km}
= e_a^i e_b^l \hat R^{ab}_{\ \ km}$ is the usual Riemann curvature tensor. 
Furthermore, we  will use 
covariant derivatives with respect to  proper time 
along a trajectory in the form 
\begin{eqnarray*}
\De P_m &=& \frac{\de P_m }{\de \tau} - \Gamma^i_{mk}P_i u^k \\
\De P_a &=& \frac{\de P_a}{\de \tau} - \Gamma^b_{\ ak} P_b u^k. 
\end{eqnarray*}
It is easily shown that $\De P_m = e^a_m \De P_a $, which justifies the 
use of the same symbol $\De $. Of course, the same derivative 
when referring to the torsion-less connection will be denoted by $\hat \De $. 

For the sake of completeness, and to avoid confusion concerning  
sign conventions, let us also derive the 
conservation laws for the stress-energy tensor and the intrinsic spin 
density (for details, see \cite{14}). The  stress-energy and 
spin density are defined by variation of the matter Lagrangian density 
$\mathcal L_m$ as follows
\begin{eqnarray}
T^a_{\ i} &=& \frac{1}{2e}\ \frac{\delta \mathcal L_m}{\delta e^i_a} \\
\sigma_{ab}^{\ \ i} &=& \frac{1}{e}\ \frac{\delta \mathcal L_m}{\delta
  \Gamma^{ab}_{\ \ i}}. 
\end{eqnarray}
Under an infinitesimal Lorentz transformation (13), with $\Lambda^a_{\ b} = 
\delta^a_b + \epsilon^a_{\ b}$, where $\epsilon^{ab} = - \epsilon^{ba}$, 
 the fields transform as 
\begin{equation}
\delta \Gamma^{ab}_{\ \ m} = - \epsilon^{ab}_{\ \ ,m}- \Gamma^a_{\ cm}
\epsilon^{cb} - \Gamma^b_{\ cm}\epsilon^{ac},\ \ \delta e^m_a = 
\epsilon_a^{\ c} e^m_c.
\end{equation}
Therefore, the variation of the matter Lagrangian reads, omitting 
a total divergence 
\begin{equation}
\delta \mathcal L_m = \frac{\delta \mathcal L_m}{\delta e^m_a} \delta e^m_a
+ \frac{\delta \mathcal L_m}{\delta \Gamma^{ab}_{\ \ m}}\delta 
\Gamma^{ab}_{\  \ m} = e(2 T^{[ac]} + \De_m \sigma^{acm}) \epsilon_{ac},
\end{equation}
where $\De_m $ is defined to act with $\Gamma^{ab}_{\ \ m}$ on tangent 
space indices, and with $\hat \Gamma^l_{ki}$ (torsion free) on 
spacetime indices. The requirement of Lorentz invariance, i.e. $\delta
\mathcal L_m = 0$, then leads to 
\begin{equation}
2 T^{[ac]}+ \De_m \sigma^{acm} = 0. 
\end{equation}
On the other hand, the connection $\Gamma^{ab}_{\ \ m}$ and the 
tetrad field $e^a_m$ are covariant 
spacetime vectors with respect to $m$ 
(thus, $e^m_a$ is a contravariant vector), and under an infinitesimal 
coordinate transformation 
\begin{equation}
\tilde x^i = x^i + \xi^i, 
\end{equation}
the fields transform as 
\begin{equation}
\tilde e^m_a(\tilde x) = e^m_a(x) + \xi^m_{,k}e^k_a,\ \ \tilde 
\Gamma^{ab}_{\  \ m}(\tilde x)=   \Gamma^{ab}_{\ \ m}(x)  
- \xi^k_{\ ,m}\Gamma^{ab}_{\ \ k}.
\end{equation}
We are interested in the change of the Lagrangian under active
transformations. In order to evaluate the change of the fields 
at the same point $x$, we have to express the transformed fields 
with the old coordinates, i.e. 
\begin{equation}
\tilde e^m_a(x) = \tilde e^m_a(\tilde x) - \tilde e^m_{a,k}(\tilde x) \xi^k, 
\ \ \tilde \Gamma^{ab}_{\ \ m}(x) = \tilde \Gamma^{ab}_{\ \ m}(\tilde x) 
- \xi^k \Gamma^{ab}_{\ \ m,k}(\tilde x). 
\end{equation}
We then find for the variation, to first order in $\xi$ 
\begin{eqnarray}
\delta e^m_a 
&=& \tilde e^m_a (x) - e^m_a(x) = \xi^m_{,k}e^k_a- \xi^k e^m_{a,k},\\ 
\delta \Gamma^{ab}_{\ \ m} &=& \tilde \Gamma^{ab}_{\ \ m}(x)
- \Gamma^{ab}_{\ \ m}(x) = - \xi^k_{\ ,m} \Gamma^{ab}_{\ \ k} -\xi^k 
\Gamma^{ab}_{\ \ m,k}. 
\end{eqnarray}
The change in the Lagrangian therefore reads 
\begin{eqnarray}
 \delta \mathcal L_m &=& \frac{\delta L_m}{\delta e^m_a} \delta e^m_a 
+ \frac{\delta \mathcal L_m}{\delta \Gamma^{ab}_{\ \ m}} \delta 
\Gamma^{ab}_{\  \ m} \nonumber \\
 &=& -2(eT^a_{\ m}e^k_a)_{,k}\xi^m - 2eT^a_{\ m}e^m_{a,k}\xi^k
+ (e\sigma_{ab}^{\ \ m}\Gamma^{ab}_{\ \ k})_{,m} \xi^k - e\sigma_{ab}^{\ \ m}
\Gamma^{ab}_{\ \ m,k} \xi^k. 
\end{eqnarray}

Requiring $\delta L_m = 0$ and regrouping carefully the terms, we finally 
get 
\begin{equation}
(\De_m T^m_{\ \ b}) e^b_k + T^m_{\ \ b} T^b_{\ mk} = 
\frac{1}{2}R^{ab}_{\ \ mk} 
\sigma_{ab}^{\ \ m} + \frac{1}{2}\Gamma^{ab}_{\ \ k} (\De_m\sigma_{ab}^{\ \ m}
+ 2 T_{[ab]}). 
\end{equation}
The last term vanishes with (18). The first term, written 
in terms of the usual (torsionless) covariant derivatives, (denoted 
with a semicolon ; ) reads  
\begin{eqnarray*}
\De_m T^m_{\ \ b}e^b_k &=& T^m_{\ \ b;m}e^b_k-K^a_{\ bm}T^m_{\ \ a}e^b_k \\ 
&=& T^m_{\ k;m}-K^l_{\ km}T^m_{\ l},
\end{eqnarray*}
and the second term, using the relation $T^i_{\ lm} =-2K^i_{\ [lm]}$, 
\begin{displaymath}
T^m_{\ \ b}T^b_{\ mk}=-T^m_{\ l}(K^l_{\ mk}-K^l_{\ km}).
\end{displaymath}
Therefore, the first two terms of (25) simplify to 
$T^m_{\ k;m}-T^m_{\ l}K^l_{\ mk}$ and the equation of motion finally takes 
the form
\begin{equation}
T^{m}_{\ k;m} - K^{im}_{\ \ k} T_{mi} = \frac{1}{2}R^{ab}_{\ \
  mk}\sigma_{ab}^{\ \ m}. 
\end{equation}
Recall that $K^{im}_{\ \ k}$ is antisymmetric in $im$, so that 
classical matter (with $T_{[mi]}= 0$ and $\sigma_{ab}^{\ \ m} = 0$) 
will follow the general relativity relation $T^{km}_{\ \ ;m} = 0$. 

Let us now return to the classical pole dipole particle 
without intrinsic spin. 
We will  re-derive the equations of the previous section 
 using the concepts of Riemann-Cartan geometry.  

Based on the form of equation (8), we will try the following Lagrangian
\begin{equation}
L = e^a_i P_a u^i - \frac{1}{2} \hat \Gamma^{ab}_{\ \ m} S_{ab} u^m. 
\end{equation}
Of course, we have coupled the (classical) spin to the torsion-less 
connection only. In (27), the momentum vector $P_a$ and the 
spin tensor $S_{ab}$ are considered as parameters and are neither functions 
of the coordinates nor of the velocities. They can be considered as 
the charges of the corresponding gauge fields $e^a_m$ 
and $\hat\Gamma^{ab}_{\ \ m}$.
  
The Euler-Lagrange equations are then  derived in a straightforward manner:
\begin{equation}
e^a_{i,m}u^iP_a - e^a_{m,i}u^iP_a -\frac{1}{2}(\hat \Gamma^{ab}_{\ \ i,m}
- \hat \Gamma^{ab}_{\ \ m,i}) S_{ab} u^i = 
e^a_m \dot P_a - \frac{1}{2} 
\Gamma^{ab}_m \dot S_{ab}. 
\end{equation}
We use the same trick as before and express $\dot P_a $ with the help 
of $\De P_a$ and $\dot S_{ab}$ with $\hat \De S_{ab}$. The result 
is
\begin{equation}
 \De P_m - T^a_{\ mi}u^i P_a + K^a_{\ im}u^i P_a = - \frac{1}{2} 
\hat R^{ab}_{\ \ mi} S_{ab}u^i 
 + \frac{1}{2} \hat \Gamma^{ab}_{\ \ m}
[\hat \De S_{ab} - u_b P_a + u_a P_b].
\end{equation}
It is easily shown, using 
$\De P_m = \hat \De P_m - K^l_{\ mi} P_l u^i$ and 
$T^l_{\ mi}= -2 K^l_{\ [mi]}$, 
that the left hand side is just $\hat \De P_m $. 

In order for the equation to be Lorentz gauge invariant, we 
have to require 
\begin{equation}
\hat D S_{ik} = u_k P_i - u_i P_k, 
\end{equation}
and then the  momentum equation reduces to 
\begin{equation}
\hat \De P_m = - \frac{1}{2} R^{lk}_{\ \ im} S_{lk} u^m. 
\end{equation}
If we define the mass through $P_i u^i = m$ and require $u_iu^i = 1$, 
we can derive from (30) the relation
\begin{equation}
P_i = m u_i + \hat \De S_{ik} u^k. 
\end{equation}
Equations (30) and (31), together with (32) are exactly the 
equations derived by Papapetrou in \cite{2}. 
The equations derived in the previous paragraph differ by the 
second term in (32), which, as we have said, is of order $S^2$. 

The derivation used in this paragraph shows clearly that the same 
equations are valid in every type of Riemann-Cartan spacetime. 
In the final equations, there is no torsion involved. The same 
equations hold, for instance, in a teleparallel theory. Torsion 
effects will only arise if we consider particles with intrinsic spin.

\section{Particles with intrinsic spin} 
  
Particles with intrinsic spin require a more careful treatment, because 
one can easily run into problems, as has been clarified in \cite{4}. 
In this section, we have in mind elementary particles (extended 
bodies with intrinsic spin will be treated in the next section). 
Therefore, it seems, at first sight, obvious, to treat them 
as monopole particles, without dipole and higher order moments. 
Let us briefly review  the Papapetrou method, as it has been 
applied in a Riemann-Cartan framework in \cite{4} and \cite{5}, 
neglecting higher order poles.  

We start from equations (18) and (26). It is convenient to 
write them with spacetime indices only, and to use the following form 
\begin{eqnarray}
(e \sigma^{ikm})_{,m} = -2 eT^{[ik]} - \Gamma^i_{lm} e\sigma^{lkm} 
- \Gamma^k_{lm} e \sigma^{ilm} \\
(e T^m_{\ k})_{,m} - \hat \Gamma^l_{km} eT^m_{\ l} - K^{im}_{\ \ k}
eT_{mi} = \frac{1}{2} eR^{il}_{\ \ mk}\sigma_{il}^{\ \ m}, 
\end{eqnarray}
where $e = \det e^a_m = \sqrt{-g}$. Then, one considers a worldline 
$X^m$ and develops all the fields around this worldline, $x^m = X^m + 
\delta x^m$. Apart from  (33) and (34), consider the following equations
\begin{eqnarray}
(e\sigma^{ikm}x^n)_{,m} &=& e \sigma^{ikn} + x^n (e\sigma^{ikm})_{,m} \\
(e T^m_{\ k}x^i)_{,m} &=& e T^i_{\ k} + x^i(e T^m_{\ k})_{,m}. 
\end{eqnarray}
Integrating over  three dimensional space and neglecting integrals 
containing  $\delta x^m$ 
(monopole approximation), one derives the following equations from (35) and 
(36)
(for details, see \cite{4}):
\begin{eqnarray}
u^i P_k = u^0 \int e T^i_{\ k}\de^3 x , \ \ \text{with} \ \ 
P_k = \int eT^0_{\ k} \de^3 x,\\
\sigma^{lm} u^k = u^0 \int e \sigma^{lmk}\de^3 x, \ \  \text{with}\ \ 
\sigma^{ik} = \int e \sigma^{ik0}\de^3 x, 
\end{eqnarray}
where $u^i = \de X^i /\de \tau$. It is easy to show that $\sigma^{ik}$ and 
$P_k$ as defined above are tensors in the monopole approximation. 
Using those relations in (33) and (34), we arrive at the following 
equations 
\begin{eqnarray}
\De \sigma^{ik} &=& (P^i u^k - P^k u^i) \\ 
\hat \De P_k - K^{im}_{\ \ k} u_m P_i &=&
 \frac{1}{2} R^{il}_{\ \ mk}\sigma_{il}u^m.
\end{eqnarray}

This completes the Papapetrou analysis of the monopole particle with 
intrinsic spin. However, as has been clarified in \cite{4}, 
there are problems with the above approach. Indeed, the relation 
(38) cannot be considered to be generally valid.  
This  
is easily seen by considering the Dirac particle, where $\sigma^{ikl}$ is 
totally antisymmetric. For such a spin density, the relation (38) leads 
to $\sigma^{ik} = 0$, as can be seen by setting one index to zero 
and using the antisymmetry properties of $\sigma^{ikl}$. 
 
Applying  WKB methods on Lagrangians for  
particles with integer 
and half-integer spins, the following relation
that replaces (38) was derived in \cite{18}
\begin{equation}
  u^0 \int e \sigma^{lmk} \de^3 x = \sigma^{lm} u^k + \frac{1}{2s}
[\sigma^{kl}u^m + \sigma^{mk}u^l],   
\end{equation}
for a particle with spin $s$. This reduces to (38) in the limit of large spin, 
which coincides with the usual result for the Weyssenhoff spin fluid. 
Thus, (38) can be considered to hold for macroscopic spin polarized bodies,
whereas elementary particles obey (41).  
For spin one-half, we get from (41) a totally antisymmetric spin density. 

In order to avoid the relation (38), that arises in the monopole approximation 
of the Papapetrou approach, 
the authors of \cite{4} concluded 
that one has to include dipole moments of the particle, i.e. to introduce 
a classical spin besides the intrinsic spin. The same has been done in 
\cite{5}, where the Mathisson-Papapetrou equations in 
Riemann-Cartan space were 
derived for the first time. This seems a rather strange 
concept for an elementary particle and is problematic from a practical 
point of view. How can one discern experimentally the spin and the 
rotational momentum of an elementary particle? 
How can we define the intrinsic spin of an 
electron if it is not 
 the total angular momentum in the particle's rest frame? 
Therefore, throughout 
this article, we take the point of view that the intrinsic spin 
of an elementary particle corresponds to its total angular momentum 
in the rest frame of the particle. There are possible objections 
to this, especially when one considers compound particles, like the 
proton for instance. In such cases, one could say that the spin 
is partly due to the spin of the constituent quarks and partly of 
orbital nature (some kind of rotation of the quarks around the center 
of mass of the proton, in a semiclassical picture). Indeed, experiments 
are carried out that aim at determining these different parts.   
However, the fact that the total spin appears always to be exactly 1/2 
leads us to believe that the proton as a whole can be treated quantum 
mechanically as a spin 1/2 particle. If the spin is partly of non-intrinsic 
nature, one has to answer the question why the orbital part cannot be 
transferred to other particles and why it always sums up with the intrinsic 
part to exactly 1/2. The final answer to this question, however, should 
belong to the experiment. To this aim, the full equations, containing 
intrinsic spin as well as dipole correction terms, as 
derived in  \cite{4} and \cite{5}, could eventually be used. 

In order to avoid the problem with equation (38) without taking into account 
dipole terms, we will choose  another way of dealing with (semiclassical) 
elementary particles. 
Having in mind that the connection $\Gamma^{ab}_{\ \ m}= \hat 
\Gamma^{ab}_{\ \ m} + K^{ab}_{\ \ m}$ couples 
linearly to the spin density in elementary particle Lagrangians,  
we conclude from the fact that  
the spin density is of the  form (41), that the spin-torsion coupling 
will be  of the form 
\begin{equation}
K^{ik}_{\ \ l}[\sigma_{ik}u^l + \frac{1}{2s}\sigma^l_{\ i} u_k + 
\frac{1}{2s}\sigma_k^{\ l} u_i] = K^{*ik}_{\ \ \ l}\sigma_{ik}u^l. 
\end{equation}
The left hand side determines $K^{*ik}_{\ \ \ l}$ to which we will refer 
as the effective torsion 
\begin{equation}
K^{*ik}_{\ \ \ l} = K^{ik}_{\ \ l} + \frac{1}{2s}K_l^{\ ik} + \frac{1}{2s}
K^{k\ i}_{\ l}.
\end{equation}
Only this part of the torsion will couple to the spin. For a macroscopic 
spin-polarized body, $K^{*ik}_{\ \ \ l}$ reduces to the full $K^{ik}_{\ \ l}$ 
and for the Dirac particle ($s= 1/2$), it reduces to the totally antisymmetric 
part of the torsion. All quantities formed from the effective torsion 
will be denoted with a star, especially $\Gamma^{*ab}_{\ \ m} = 
\hat \Gamma^{ab}_{\ \ m} + K^{*ab}_{\ \ \ m}$. 

Let us now derive the equations of motion for spin and position 
for a point particle with 
intrinsic spin. If we suppose that in a purely Riemannian space, 
the particle with intrinsic spin will behave just like a particle 
with classical spin, we have to introduce the term $-\frac{1}{2}
\hat \Gamma^{ab}_{\ \ m}\sigma_{ab} u^m$ into our Lagrangian (see (27)).
If we add to this  the spin-torsion coupling 
(42), we are led to the following Lagrangian 
\begin{equation}
L = e^a_i P_a u^i - \frac{1}{2} \Gamma^{*ab}_{\ \ m} \sigma_{ab} u^m. 
\end{equation}
The ultimate physical justification for the use of a specific 
Lagrangian can only 
be the correctness of the equations of motion derived from it. 

Just as in the previous sections, we readily derive the Euler-Lagrange 
equations in the form 
\begin{equation}
 \De^* P_m - T^{*a}_{\ \ mi}u^i P_a =  
- \frac{1}{2} R^{*ab}_{\ \ mi}\sigma_{ab}u^i   
+ \frac{1}{2} \Gamma^{*ab}_{\ \ m} [ \De^* \sigma_{ab} - P_a u_b + P_b u_a]. 
\end{equation}
Requiring the gauge invariance under a local Lorentz transformation, we 
finally get 
\begin{eqnarray}
\De^* P_m - T^{*a}_{\ mi}u^i P_a &=&
- \frac{1}{2} R^{*ab}_{\ \ mi}\sigma_{ab}u^i, \\  
 \De^* \sigma_{ik} &=& P_i u_k - P_k u_i. 
 \end{eqnarray}
In order to compare with the classical geodesics, we can rewrite the first 
equation as 
\begin{equation}
\hat \De P_m - \frac{1}{2}K^*_{lim}(P^lu^i- P^iu^l) = - \frac{1}{2}
 R^{*ab}_{\ \ mi}\sigma_{ab}u^i . 
\end{equation}
For a macroscopic spin-polarized body, these equations agree completely with 
the result from the monopole approximation in \cite{4}, i.e. with 
equations (39) and (40). However, such a macroscopic body will 
also possess dipole moments and the equations have to be adopted 
to that case (see next section). 
For the 
spin half particle (totally antisymmetric torsion), the results agree, 
if we ignore again the quantity $P^{[i}u^{k]}$, which is of order 
$\sigma^2$, with the results from the WKB analysis of the Dirac equation, 
as carried out in \cite{19}. Also, in the same approximation, there is 
an agreement with the results of \cite{20} for the spin 3/2 particle 
and of \cite{21} and \cite{22} for the spin 1 or Proca particle.  

Thus, there is no need to consider dipole moments while dealing with 
elementary particles. The Lagrangian (44) provides a very comfortable 
method of deriving the correct equations of motion without reference neither 
to the gravitational field equations, nor to the specific matter Lagrangian 
for the particle under consideration. The only additional information we 
needed was relation (41). 

Equations (46) and (47) were derived in \cite{7} for the case 
of a purely Riemannian space ($T^a_{\ lm} = 0$), following a 
slightly different procedure. The term $\Gamma^{ab}_{\ \ m} \sigma_{ab}u^m$
was interpreted as the Hamiltonian part of a Routhian $R$ and the spin 
evolution equations
were then derived using the Heisenberg equations 
$\frac{\de \sigma_{ab}}{\de \tau}= i[ R, \sigma_{ab}]$ and the  
commutation relations for the spin tensor. However, in 
order to get the correct 
equations, they had  to redefine the spin tensor and to modify 
slightly the Routhian. This was due to the fact that their Lagrangian 
part was taken to be of the form $m\sqrt{u_iu^i}$. On the other hand, 
if we  start with 
the form $e^a_i P_au^i$ and suppose that $P_a$ and $\sigma_{ab}$ 
obey the commutation relations of the Poincar\'e algebra, one 
 gets the correct precession equation following the same 
procedure as in \cite{7}.   
We can see this as an independent confirmation of our method, since 
the mere requirement of Lorentz gauge invariance leads to the same result. 

In \cite{8}, a different method was used, involving additional variables 
to describe the spin degrees of freedom. However, although initially intended 
to describe elementary particles in Riemann-Cartan spacetimes, the 
equations actually correspond to the special case $s \rightarrow \infty$. 
The same holds true for the equations given in \cite{23} and \cite{24}. 
Another, more recent attempt to 
generalize the multipole formalism to Riemann-Cartan 
geometry can be found in \cite{25}.

\section{Macroscopic bodies with intrinsic spin}

Finally, we are ready to consider macroscopic spin-polarized bodies. 
This is probably the experimentally most important case, since gravitational 
effects on elementary particles are usually very small. (See however 
\cite{26}-\cite{28} for experiments involving gravity on a quantum mechanical 
scale.)
In this section, 
we have in mind an extended body, like a neutron star, for which the spins 
of the constituent particles are (partially or fully) aligned. In
addition, the body may rotate. From  the last two sections, 
we may expect a coupling of the classical spin $S_{ab}$ to the Riemannian 
connection $\hat\Gamma^{ab}_{\ \ m}$, 
and a coupling of the intrinsic spin $\sigma_{ab}$ to the effective 
connection, which coincides in the macroscopic case with the full 
connection $\Gamma^{*ab}_{\ \ m}= \Gamma^{ab}_{\ \ m}$. Hence we start with
\begin{equation}
L = e^a_i P_a u^i 
- \frac{1}{2} \hat\Gamma^{ab}_{\ \ m} S_{ab} u^m
- \frac{1}{2} \Gamma^{ab}_{\ \ m} \sigma_{ab} u^m.
\end{equation}
We derive the following equations
\begin{eqnarray}
 \hat \De P_m - \frac{1}{2} K_{lim}(P^lu^i-P^iu^l) &=& 
- \frac{1}{2}\hat R^{lk}_{\ \ mi}S_{lk}u^i   
- \frac{1}{2} R^{ab}_{\ \ mi}\sigma_{ab}u^i  \nonumber \\ 
&&+  \frac{1}{2} \Gamma^{ab}_{\ \ m} [\De \sigma_{ab} - P_au_b + P_b u_a] 
+ \frac{1}{2} \hat \Gamma^{ab}_{\ \ m} \hat \De S_{ab}.
\end{eqnarray}
We can now regroup all the terms proportional to $\Gamma^{ab}_{\ \ m}$,  
or equivalently to $\hat \Gamma^{ab}_{\ \ m}$, and then require 
Lorentz covariance. The result is 
\begin{equation}
 \hat \De P_m - \frac{1}{2} K_{lim}(P^lu^i-P^iu^l) = 
- \frac{1}{2}\hat R^{lk}_{\ \ mi}S_{lk}u^i   
- \frac{1}{2} R^{ab}_{\ \ mi}\sigma_{ab}u^i   
- \frac{1}{2} K^{ik}_{\ \ m} \hat \De S_{ik}
\end{equation}
and for the spin evolution
\begin{equation}
\De \sigma_{ik} + \hat \De S_{ik} = P_i u_k - P_k u_i. 
\end{equation}
We can use (52) in (51) to eliminate the apparent coupling of the 
torsion to the rotational moment in the last term. 
The equations are more or less what could have been expected right from the 
start, except maybe for the last term in (51). 
Our  equations agree with those derived in  \cite{4} 
(for the macroscopic limit $s \rightarrow \infty$) using 
the Papapetrou method in the dipole approximation.  

These equations alone are of course not sufficient to determine 
the behavior of the body. An additional relation between the 
internal and the classical spin has to be assumed. For a simple model  
of a neutron star, let us suppose  a strong coupling of the form 
$S_{ik} = g \sigma_{ik}$ with constant $g$. In this case, we can write 
(52) in the form 
\begin{displaymath}
\frac{1+g}{g} \left[ \hat \De S_{ik} + \frac{1}{1+g}K^l_{\ im} S_{lk} u^m 
+ \frac{1}{1+g}K^l_{\ km} S_{il} u^m \right]=
 P_i u_k - P_k u_i. 
\end{displaymath}
Now, if we introduce the connection 
\begin{equation}
\tilde \Gamma^i_{lm} = \hat \Gamma^i_{lm} + \frac{1}{1+ g}\ K^i_{\ lm}, 
\end{equation}
we can write 
\begin{displaymath}
 \frac{1+g}{g}\  \tilde \De S_{ik} = P_i u_k - P_k u_i. 
\end{displaymath}
Multiplying with $u^i$, defining the mass as $m = P_i u^i $ and using 
$1 = u_iu^i$, we find 
\begin{equation}
P_k = m u_k + \frac{1+g}{g}\ \tilde \De S_{ki}\ u^i, 
\end{equation}
and the precession equation finally reads 
\begin{equation}
\tilde \De S_{ik} = (u_k \tilde \De S_{il} - u_i \tilde \De S_{kl}) u^l. 
\end{equation}
Thus, the intrinsic spin (and also the classical as well as the total 
spin) is Fermi-Walker transported with respect to the connection (53). 

Introducing the constraint $S_{ik}= g\sigma_{ik}$ 
directly into (49), we see that the connection 
(32) governs the whole propagation, and not only the spin evolution. The  
 equation for momentum propagation is easily derived then. 

We used the constraint $S_{ik} = g\sigma_{ik}$ mainly as 
an illustrative example. For realistic neutron star models, you 
can consult \cite{29} and \cite{30} for instance. You should however 
have in mind that these models rely strongly on general relativity 
as underlying gravitational theory. Since the use of our equations 
of motion naturally suppose that the underlying theory is a Poincar\'e 
gauge theory with dynamical torsion fields, these models have to be 
revisited in view of the modified gravitational interaction. This might lead 
to severe changes, since the interior of neutron stars is governed by 
very strong fields. To derive such models is beyond the scope of this
article. 

\section{The Dirac particle} 

In order to gain more confidence in the procedure we followed in 
section 4, mainly the in introduction of the effective torsion (43), 
we will have a closer look at the most important example, that of 
a spin 1/2 particle and present an alternative derivation of 
the equations of motion. 

The Dirac particle is described by the following Lagrangian density
\begin{equation}
\mathcal L = \frac{i}{2}\left[ \bar \psi \gamma^m D_m \psi - \bar D_m 
\bar \psi \gamma^m \psi \right ] + m\bar\psi \psi, 
\end{equation}
where 
$D_m \psi = (\partial_m - \frac{i}{2}\Gamma^{ab}_{\ \ m}\sigma_{ab})\psi, 
\  \bar D_m \bar\psi = \partial \bar \psi + \frac{i}{2}\Gamma^{ab}_m \bar\psi 
\sigma_{ab}$ 
and $\gamma^m = e_a^m \gamma^a$, with $\gamma^a $ the usual Dirac 
matrices and $\sigma_{ab} = \frac{i}{4}[\gamma_a, \gamma_b]$. 
The Dirac equation reads ($T_m = T^i_{\ mi}$)
\begin{equation}
i(\gamma^m D_m - \gamma^m T_m ) \psi = m \psi. 
\end{equation}

The stress-energy tensor $(2e)^{-1}\ \delta(e\mathcal L)/ \delta e_a^i $ 
is found to be  
\begin{equation}
T^a_{\ i} = \frac{i}{4}[\bar \psi \gamma^a D_i \psi - \bar D_i \bar \psi
\gamma^a \psi ]. 
\end{equation}
Several remarks are in order at this point. It is a well known fact 
that only the totally antisymmetric part of the torsion couples to the 
Dirac particle. This follows immediately from (57), which can be written as
\begin{equation}
i \gamma^m D^*_m  \psi = m \psi,  
\end{equation}
where $D^*_m $ is constructed from the connection (43) with $s =1/2$.  
Furthermore, the Lagrangian (56) is numerically equal to 
\begin{equation}
\mathcal L^* = \frac{i}{2}\left[ \bar \psi \gamma^m D^*_m \psi - \bar D^*_m 
\bar \psi \gamma^m \psi \right ] + m\bar\psi \psi.   
\end{equation}
This Lagrangian is preferable from the point of view that it incorporates 
a minimal coupling principle that can be applied equally well in the 
Lagrangian or in the field equations. However, (56) and (60), although 
they both lead to the same Dirac equation, are not completely 
equivalent, since for 
the latter the stress-energy tensor becomes  
\begin{equation}
T^{*a}_{\ \ i} 
= \frac{i}{4}[\bar \psi \gamma^a D^*_i \psi - \bar D^*_i \bar \psi
\gamma^a \psi ]. 
\end{equation}
which differs from (58) in the non-axial torsion parts contained in 
the latter. Once again (61) seems preferable, since the 
Dirac particle does not couple to these torsion parts. However, 
it makes of course a difference which of these tensors actually 
represents the source term of the gravitational equations. If we 
remain in the logic of a Lorentz gauge theory and have in mind 
that there might be additional sources with different spin (coupling 
to other torsion parts), the only 
consistent way is to write down only Lagrangians that contain the full 
connection $\Gamma^{ab}_{\ \ m}$, as is the case for (56), even though 
they might contain parts which do not couple to the particle in question. 
Otherwise we 
can not carry out the variation with respect to $\Gamma^{ab}_{\ \ m}$.  

However, as long as we are interested only in the equations of motion 
for the Dirac particle (or their semiclassical limit), 
and not in the gravitational field equations, 
the use of either $\Gamma^{ab}_{\ \ m}$ or $\Gamma^{*ab}_{\ \ \ m}$ is 
equivalent. Equations (57) and (59) are exactly the same. Under this 
aspect, the use of $\Gamma^{*ab}_{\ \ \ m}$ is certainly preferable.  
It can be  seen as the physically relevant  
connection since it is the  connection that really 
couples to the Dirac particle.

What does this have to do with our equations of motion? Well, our 
point is that, if we start, as in the usual Papapetrou procedure, 
from the divergence relations for the stress-energy tensor and 
the spin density, and then define the momentum as 
$P_i = \int e T^0_{\ i}\de^3 x$, 
this is already very unfortunate right from the start, since this 
momentum vector contains implicitly non-axial torsion parts from (58) 
which should not couple to the Dirac particle at all. This 
explains at least partly why the monopole approximation does not 
lead to correct results for elementary particles \cite{4}. A first 
step should be to look for a divergence relation involving the 
tensor $T^{*a}_{\ \ i}$ and carry out the Papapetrou method on this 
tensor.  

The above considerations are also useful  in another aspect. 
The use of the covariant derivative $D^*_m$ allows us to 
generalize the proper time formalism, i.e. the covariant generalization of 
the 
Heisenberg equations,   
 to the case of a Riemann-Cartan space. 
We will briefly discuss the concepts and refer to \cite{31}-\cite{34}  
and the references therein for further details. Notably, in \cite{34}, 
the resulting equations were used to extract for the first time 
the precession frequency of the spin of a Dirac particle in 
an axial torsion field. 
 
One  starts with the operator $H$ whose eigenvalue is  (the negative of)
 the mass, $H\psi = - m\psi$. 
(The mass is the only available covariant 
generalization of the classical energy, apart from the 
stress-energy tensor.) Then identify (for any operator $A$) 
\begin{equation}
i[H, A] = \frac{\de A}{\de \tau},  
\end{equation}
where $\tau$ is interpreted as  proper time along 
the semiclassical trajectory. 
In flat spacetime, the Hamiltonian is just $H = - \gamma^i p_i $, 
with $p_i = i \partial _i $ and EM fields can easily be taken into 
account by $p_i \rightarrow \pi_i = i \partial_i + eA_i$. With  
this in mind, and having a look at (59), we find that the Hamiltonian for our 
case has  the form 
\begin{equation}
H = - \gamma^m P_m, 
\end{equation}
with 
\begin{equation}
P_m =  i D^*_m = i (\partial_m - \frac{i}{2}\Gamma^{ab*}_{\ m}\sigma_{ab}),
\end{equation}
where $\Gamma^{ab*}_{\ \ m}$ is again the connection which contains only 
the totally antisymmetric part of the torsion. For the velocity operator, 
we find 
\begin{equation}
u^i = \frac{\de x^i}{\de \tau} = i[H,x^i] = \gamma^i. 
\end{equation}
We thus have the expected relation $H = - u^mP_m  = - m$. 
Let us have a look at the Heisenberg algebra for momentum and 
position operators. We have:
\begin{displaymath}
[x^i, x^k ] = 0,\ \ \  [ P_i, x^k ] =  i \delta^k_i,  
\end{displaymath}
\begin{displaymath}
[ P_i, P_k ] = \frac{i}{2}R^{*cd}_{\ \ \ ik}\sigma_{cd}.  
\end{displaymath}
This is in complete analogy with the EM case (just recall the 
relation $[\pi_i, \pi_k] = i e F_{ik}  $). A useful relation is 
also $[P_i, \gamma^k] = - i \gamma^l \Gamma^{*k}_{\ \ li}= 
- i \gamma^l(T^{*k}_{\ \  il} + \Gamma^{*k}_{\ \ il})$.

Writing down the commutation relations for $P_m, \gamma^i $ and $\sigma_{ab}$ 
with the Hamiltonian, it is straightforward to derive the 
following equations of motion 
\begin{eqnarray}
\De^* \sigma_{lk}&=& P_l u_k  - P_k u_l  \\
\De^* P_m &=& - \frac{1}{2}R^{*cd}_{\ \ \ mk}u^k \sigma_{cd} 
+ T^{*k}_{\ \ ml}u^l P_k \\
\De^* u^i &=& -\sigma^{mi}P_m . 
\end{eqnarray}
Of course, $\De^*$ appearing here is again the proper time derivative 
acting on tensors (not on spinors) with $\Gamma^{*i}_{\ lm}$. 
Relation (68) has been related to the zitterbewegung and has no 
classical counterpart (see \cite{31}). The other two are exactly our 
equations (46)-(47). 

Both derivations, the one of section 4 and the present one, although 
conceptionally quite different, lead to the correct results of the WKB 
analysis  \cite{19}. In both cases, the crucial step was to use 
the correct connection containing only the part of the torsion that 
effectively couples to the spin density.

\section{Spin-polarized test body in Brans-Dicke theory}

Finally, we will take a look at a specific example involving 
the equations derived in section 5. Since in most practical cases, 
intrinsic 
spin effects will be very small (see however \cite{26}-\cite{28}), 
we turn to  a theory  that 
gives rise to a torsion field even in the absence of a spinning source, 
namely Brans-Dicke theory generalized to Riemann-Cartan geometry. 
The theory is based on the action 
\begin{equation}
S = \int e\ \left(\phi R - \frac{\omega + \frac{3}{2}}{\phi}\  \phi_{,m}
\phi^{,m} - \mathcal L_m\right) \de^4 x.  
\end{equation}
It is easily shown by solving the equation that arises from variation 
with respect to the connection, that 
in the case of a spinless matter Lagrangian (the source), 
we get a torsion field 
of the form (cf. \cite{35}, \cite{36}), 
\begin{equation}
K^{lm}_{\ \ i} = \frac{1}{2\phi}(\delta^l_i \phi^{,m} - \delta^m_i \phi^{,l}). 
\end{equation}
The field $\phi$ and the metric $g_{ik}$ are subject to the classical 
Brans-Dicke field equations, as can be seen by plugging (49) into the 
equations for $\phi$ and $e^a_m$. The torsion (70) is entirely 
determined by its trace alone.  
It is often claimed (probably having the 
Dirac particle in mind), that such a torsion field is pure gauge, and 
does not lead to physical consequences. However, if the spin of the 
test particle moving in this field is 
not totally antisymmetric, there  might well 
be a measurable effect. 

We are interested in a macroscopic test body with intrinsic spin moving 
in the field of a classical, spherically symmetric source. 
(For equations of motion of a classical point mass in Brans-Dicke theory, 
see for instance \cite{37} and references
therein.)
For 
simplicity, although not quite realistic, let us suppose that 
the test body does not rotate. The spin evolution is then 
described by equation 
(52) with $S_{ik} = 0$. We proceed as in \cite{4} and introduce the 
spin vector 
 $\sigma^i = \frac{1}{2}\eta^{iklm}u_k \sigma_{lm}$ where $\eta^{iklm} = 
\frac{1}{e} \epsilon^{iklm}$. If we impose the condition
$\sigma_{kl}u^k= 0$ we can invert to $\sigma^{ik} = - \eta^{iklm}u_l
\sigma_m$. 
Omitting  the higher order terms, we are 
left with 
\begin{equation}
\De \sigma^i = \hat \De \sigma^i  + K^i_{\ lm}\sigma^l u^m = 0.  
\end{equation}
 We are not interested in the momentum equation, because 
the deviations from the geodesics are expected to be very small. 
  
Since the metric and scalar field are determined from the same equations 
as in classical Brans-Dicke theory, we can directly use the results 
from the post-Newtonian expansion as given in \cite{38} for instance, 
and write 
\begin{eqnarray}
\phi &=& 1 + c \frac{m}{r} \\  
g_{00} &=& 1 - 2 a \frac{m}{r} \\
g_{\alpha\beta} &=& \delta_{\alpha\beta} ( -1-2b\frac{m}{r}). 
\end{eqnarray}
The coefficients read 
\begin{eqnarray}
a &=& 1 - \frac{s}{2+ \omega} \\ 
b &=& \frac{1+ \omega}{2+ \omega}(1+ \frac{s}{1+ \omega}) \\
c &=&\frac{1-2s}{2+ \omega}.  
\end{eqnarray}
Here, $m$ is the mass of the source body and $s$ its sensitivity 
(see \cite{38} and references therein). The sensitivity is basically 
the binding energy per unit mass. Recall that the static spherically 
symmetric solution of Brans-Dicke theory is not a one parameter solution 
as in general relativity, but depends on more  properties of 
the central source. In the parameterized post-Newtonian formalism, it 
turned out that an astrophysical body can conveniently be described 
by its mass and its sensitivity. The latter takes values from 
$10^{-6}$ for a usual star (like the sun) to one half for black holes. 
Neutron stars have values around $0.1-0.3$ (see \cite{38}). 

We now introduce (72)-(74) into (70) and (71). From the construction of 
$\sigma^i$, 
we have $\sigma^iu_i = 0$. Using this relation and parameterizing the 
equation with  the time coordinate, we get the result 
\begin{eqnarray}
 \frac{\de \vec \sigma}{\de t} &=& \frac{m}{r^3}  
\left [    \frac{a+2b+c}{2}
(\vec L \times
\vec \sigma ) 
\right. \nonumber\\  
&& \left. 
- \frac{a}{2}(\vec x (\vec \sigma \cdot \vec v) + \vec v (\vec \sigma 
\cdot \vec x))
+ b \vec \sigma ( \vec x \cdot \vec v)  
+ \frac{c}{4}(\vec v ( \vec x \cdot \vec \sigma) + \vec x(\vec v \cdot \vec 
\sigma)) 
\right ] , 
\end{eqnarray}
with $\vec L = \vec v \times \vec x$ the orbital angular momentum per 
unit mass of the test body. 
We have separated terms antisymmetric and symmetric 
in $(\vec x, \vec v)$. 
Just as in general relativity, 
the latter can easily be shown to 
have  mean values over one orbit that  
vanish for closed orbits and that are  
 negligible for the quasi-closed 
orbits we consider here (i.e. orbits closed, up to a small spin precession 
correction). 

Thus, the relevant  precession equation reads:
\begin{equation}
 \frac{\de \vec \sigma}{\de t} = \frac{m}{r^3} \  \frac{a+2b+c}{2}\  \left(
\vec L \times
\vec \sigma \right) . 
\end{equation}
The torsion corrections come from the term in $c$. 

In the limit $\omega \rightarrow
\infty$,  the parameters take the values $a= 1, b= 1, c = 0$ and 
we get the well known factor $\frac{3}{2}$ from general relativity 
in the spin precession 
equation, but this time for intrinsic spin. This confirms once 
again the result that in general relativity, a particle with 
intrinsic spin behaves just like a particle with classical spin. 
More precisely, in the $\omega \rightarrow \infty$ limit, the theory 
goes over not to general relativity, but to Einstein-Cartan theory (see 
\cite{39} for a recent review). 
Since we consider only spinless sources however, there is no difference 
in the resulting solutions. (Even for a spinning source, the differences 
vanish outside the source.)

In Brans-Dicke theory, the only case where the torsion vanishes is 
when $s = \frac{1}{2}$, i.e. when the source is a black hole. In this 
case, $c= 0$ and the scalar field is constant. 
  
The interesting cases are of course when $c \neq 0$. The largest influence 
of the torsion is found for ordinary stars, when $s\approx 0$. 
We can evaluate for this case 
the ratio of the torsional part and the other parts 
for different values of $\omega$, for instance 
\begin{eqnarray*}
\omega \approx 500 &\rightarrow &
\frac{c}{a + 2b} \approx 1,4 \cdot 10^{-3}  \\
\omega \approx 1000 &\rightarrow &
\frac{c}{a + 2b} \approx 6 \cdot 10^{-4} . \\
\end{eqnarray*}
These results can be interpreted in two ways: Within the framework of 
Brans-Dicke theory with torsion, they tell us the difference in 
the precession equation of a classical rotating test body (which does not 
couple to torsion) and a spin polarized test body. Thus, we can in principle 
determine, whether a body is just rotating or possesses intrinsic spin. 

On the other hand, for a spin-polarized body, they tell us 
the difference between the precession in classical Brans-Dicke 
theory without torsion for which   the spin precession 
is the same for rotational and intrinsic spin and the precession 
of the same body in Brans-Dicke theory in a 
Riemann-Cartan geometry. 

The results are easily generalized to the more realistic case of a 
rotating test body using (55). More work is involved to get results for the 
two body problem (for instance a binary consisting of two rotating, 
spin polarized neutron stars). However, as far as the test body is 
concerned, we have shown that already the pure spin terms are very 
small. If there is an additional rotation, the relevant connection 
 will be of the form (53), where for a realistic 
neutron star model ($S_{ik} >> \sigma_{ik}$), we will have a large $g$ 
and therefore, the torsion effects will become even smaller. 

As far as the source is concerned, in a realistic 
neutron star model, the rotational 
effects will by far dominate the spin effects, so that the field 
will essentially be a Kerr-like analogue of Brans-Dicke theory. 
In any case, the torsion outside the source 
will again be produced by the scalar field only, 
because the spin of the source gives rise only to non-dynamical torsion 
fields (vanishing outside of the spin density), 
just as in Einstein-Cartan theory. A discussion of geodesics and 
autoparallels (for classical point particles) in such 
a Kerr-Brans-Dicke geometry can be found 
in \cite{40}. These results may easily be generalized to take into account 
the spin effects.

Nevertheless, the analysis shows that there are  physical 
consequences of the vector torsion, although very small, 
 and we cannot consider it as a pure gauge field. 

\section{Symmetries of the Lagrangian}

Throughout this paper, we have obtained the spin evolution equation 
by requiring the momentum equation that results from the Euler-Lagrange 
equations of the Lagrangian under investigation to be either 
covariant under spacetime diffeomorphisms in the Riemannian case   
or under local Lorentz transformations  in the Riemann-Cartan 
case. One might wonder whether the same argument can be applied  
directly to the Lagrangian, i.e. if the requirement of the invariance of 
the Lagrangian under those transformations leads to the equation for the 
spin evolution. We will investigate this subject case by case. 

Let us begin with the Lagrangian (8). The only symmetry that occurs in 
Riemannian geometry is the diffeomorphism invariance, which can be 
written in the infinitesimal form 
\begin{equation}
x^i \rightarrow x^i + \epsilon^i(x). 
\end{equation}
This contains Lorentz transformations ($\epsilon^i(x) = \epsilon_{ik}x^k$ 
with $\epsilon_{ik}$ antisymmetric and constant) as well as translations 
($\epsilon^i = a^i$ with constant $a^i$) as subcases. The requirement of 
the Lagrangian to be invariant (up to a proper time derivative) under 
(80) leads to the Euler-Lagrange equations if the Lagrangian is 
a function of $x^i$ and $u^i$ alone. However, in the  case of (8), 
there are 
additional quantities $S_{ik}$ that transform as a tensor under coordinate 
transformations. Therefore, the requirement $\delta L = 0$ leads to 
the  equation $\dot S_{ik} = 0$, as is easily shown. This 
cannot be used as a constraint;  it is neither covariant, nor  physically 
acceptable as equation of motion.  
Therefore, we can conclude that the  Lagrangian (8) is not a scalar. Covariant 
under (80) are only the final equations of motions. 

However, in the case of the Lagrangians (27), (44) and (49), no such 
problems occur, since 
the quantities $P_a$ and $S_{ab}$ transform as scalars 
under coordinate transformations. The requirement for $L$ to be invariant 
under (80) simply leads to the Euler-Lagrange equations. 
In those cases, the scalar character 
of the Lagrangians  
is obvious anyway, since the connection as well 
as the tetrad field transform as covariant spacetime vectors. 

Let us turn to the gauge transformations. We consider the Lagrangian 
\begin{equation}
L = e^a_i P_a u^i - \frac{1}{2} \Gamma^{ab}_{\ \ m} S_{ab} u^m,  
\end{equation}
which stands exemplary for the Lagrangians (27), (44) or (49). (Remember that 
the connections, $\Gamma^{ab}_{\ \ m}, \hat \Gamma^{ab}_{\ \ m}$ 
and $\Gamma^{*ab}_{\ \ m}$ all transform in the same way.) Before turning 
to the Lorentz transformations, let us make a remark on gauge translations. 

In Poincar\'e gauge theory, the ten gauge fields $\Gamma^{ab}_{\ \ m}$ 
(Lorentz) and $\Gamma^a_{\ m}$ (translations) alone are not sufficient to 
write down an invariant Lagrangian. This is due to the fact that the 
translational field $\Gamma^{a}_{\ m}$ does not transform homogeneously 
as Lorentz vector. (It might even vanish for instance and is certainly not 
invertible, and thus not suitable to be used as a tetrad field.)
One therefore introduces an additional field $\xi^a$ (the coset parameters)
transforming as $\xi^a \rightarrow \xi^a + \epsilon^a_{\ b}\xi^b + \epsilon^a$
under Poincar\'e transformations and defines the tetrad field as 
$e^a_m = \Gamma^a_{\ m} + \De_m \epsilon^a$. Then, this tetrad field 
transforms homogeneously as Lorentz vector 
(i.e. it is invariant under translations). Therefore, every Lagrangian 
expressed in terms of the fields $\Gamma^{ab}_{\ \ m}$ and $e^a_m$ is 
trivially invariant under translations. The translational gauge symmetry 
is hidden. From this point, we started our discussion of Riemann-Cartan 
geometry in section 3. The details and the foundation 
of this interpretation of  Poincar\'e 
gauge theory can be found in \cite{41}. 

This leaves us with the local Lorentz transformations. The field 
transformations as given in  section 3 can be written for 
infinitesimal Lorentz transformations with parameters $\epsilon^{ab}(x)= 
- \epsilon^{ba}$ in the form 
\begin{equation}
\delta \Gamma^{ab}_{\ \ m} = -\De_m \epsilon^{ab}\ \ \  \text{and}\ \ \  
\delta e^a_m = \epsilon^a_{\ b}e^b_m. 
\end{equation}
Let us for one moment assume that the quantities $P_a$ and $S_{ab}$ 
in (81) are invariant (i.e. scalars). Then, the change in the Lagrangian 
reads
\begin{equation}
\delta L = \epsilon^a_{\ b} e^b_m P_a u^m + \frac{1}{2}\De_m \epsilon^{ab}
 S_{ab}. 
\end{equation}
Dropping a total proper time derivative and using the antisymmetry of 
$\epsilon^{ab}$, this can be written in the form
\begin{equation}
\delta L = -\frac{1}{2} \epsilon^{ab}(\De S_{ab} - (P_au_b - P_bu_a)).
\end{equation}
Thus, the requirement $\delta L = 0$ leads to the correct spin evolution 
equation  $\De S_{ab} = P_au_b - P_bu_a$. 

However, this result was derived assuming that $S_{ab}$ and $P_a$ are 
invariant under Lorentz transformations. Since they are ultimately 
related to spacetime tensors via $S_{ik} = e_i^ae_k^bS_{ab}$ and $P_i = e_i^a
P_a$, we know that they have to transform as Lorentz tensors, as has been 
assumed throughout the article. Including this  additional variation in (83) 
leads to $\delta L = - \frac{1}{2} \epsilon^{ab} \dot S_{ab}$. If  we require 
this to vanish, we are let to  equation $\dot S_{ab} = 0$, which not only 
is not covariant, but has also to be rejected on physical grounds. 

Therefore, we conclude that the Lagrangians  (27), (44) and (49) are not 
invariant under local Lorentz gauge transformations.  
Covariant are only the final equations of motion.

\section{Conclusion}

Our main result is summarized in the Lagrangians (44) for elementary 
particles and (49) for macroscopic bodies. Varying with respect 
to the coordinates and requiring gauge or diffeomorphism 
invariance of the resulting 
equation, the equations of motion for spin and position are obtained 
very easily and without any ambiguities. 

The method can be justified by 
the fact that the results are in agreement with the WKB analysis 
of elementary particles and with those obtained from the pole-dipole 
expansion following Papapetrou's method. 

We did not discuss in general the final equations in this paper, because they 
are not new, except for the result that a rotating, spin-polarized 
body with a strong spin-rotation coupling $ S = g \sigma $, 
couples to the connection given by (53). Nevertheless, we took the 
opportunity to show briefly that the vector-torsion arising in 
generalized Brans-Dicke theory affects the spin precession of a 
macroscopic body and should not be referred to as pure gauge.  

Finally, we showed that the Lagrangians under investigation are not 
scalars under the transformations under consideration, nor does the 
requirement for them to be invariant lead to the correct spin evolution 
equations. This means that the processes of varying with respect to the 
coordinates and requiring gauge or diffeomorphism invariance do not 
commute.

\section*{Acknowledgments}

This work has been supported by EPEAEK II in the framework of ``PYTHAGORAS 
II - SUPPORT OF RESEARCH GROUPS IN UNIVERSITIES'' (funding: 75\% ESF - 25\% 
National Funds).

\end{document}